\documentclass[preprint]{aastex}

\begin{document}

\title{The CALSPEC Stars P177D and P330E}
\author{Ralph~C.\ Bohlin\altaffilmark{1} and Arlo U. Landolt\altaffilmark{2}} 
\altaffiltext{1}{Space Telescope Science Institute, 3700 San Martin Drive,
Baltimore, MD 21218, USA}
\email{bohlin@stsci.edu}
\altaffiltext{2}{Department of Physics and Astronomy Louisiana State
University Baton Rouge, Louisiana 70803}
\email{landolt@phys.lsu.edu}

\begin{abstract}

Multicolor photometric data are presented for the CALSPEC stars P177D and P330E.
Together with previously published photometry for nine other CALSPEC standards,
the photometric observations and synthetic photometry from HST/STIS
spectrophotometry agree in the B, V, R, and I bands to better than $\sim$1\% 
(10 mmag).

\end{abstract}  

\keywords{stars:individual ---
stars:fundamental parameters (absolute flux) --- techniques:photometry}

\section{INTRODUCTION}

The CALSPEC\footnote{http://www.stsci.edu/hst/observatory/crds/calspec.html}
stars are a group of spectrophometric flux standards used to calibrate
instrumentation on the Hubble Space Telescope (HST) \citet{Bohlin07},
\citet[][B14]{Bohlinetal14}. P177D is of spectral type G0 V and P330E is of
spectral type G2 V. Originally, \citet{Colina97} quoted G0 V for P330E, while
the revision to G2 V from Simbad is reasonable, based on the similarity to the
Solar spectral energy distribution (SED) as illustrated in figure 8 of B14. The
J2000 coordinates from  SIMBAD are listed in Table 1. Most of the additional
CALSPEC stars are presented in \citet{Landoltuom07,Landolterr07}, while GD71
appears in  \citet{landolt92,landolt09}. The previously unpublished GD153 is
included in Table 1 with its rms scatter, because the error-in-the-mean is not
valid with only four measures. The two G stars, P177D and P330E, are crucial
additions for cross-comparison of the Landolt and CALSPEC data, because most of
the other stars in common are of much hotter spectral types.

The photometric observations are described in Section 2, while Section 3
discusses the comparison of the actual Landolt photometry with the synthetic
CALSPEC photometry.

\section{OBSERVATIONS} 

The broad-band UBVRI photometric data for P177D and P330E were obtained during
two observing runs, separated by one year, at  the Lowell Observatory's 1.8-m
Lowell Perkins telescope. A complete data set for an observation of a star
consisted of a series of measures VBURIIRUBV. The data acquisition, reduction
and analysis have been  described in \citet{Landolt13} and also see
\citet{Landolt07}. The new photometry for P177D and P330E is tied to the
standard stars of \citet{landolt09}.
Final magnitudes and color indices for P177D and P330E are given in Table 1. 
The uncertainty (Unc) beneath each magnitude or color index is the rms mean
error for a single observation.

The individual data points that were used to derive the final magnitudes and
color indices are shown in Table 2. The UT date of observation is
given in the second column, followed by the Heliocentric Julian Date (HJD) of
the central time of each observation. The remaining columns list the magnitude
and color indices from each observation. Each star was measured twice a night on
five nights over a one year time interval, i.e. ten measures total for each
star.  Immediately beneath the final magnitude and color indices are two lines
indicating the averages and rms 1$\sigma$ errors for a single observation of
each magnitude or color index. The final line provides the mean error of the
mean, i.e. the uncertainty in the mean, for the star's magnitude or color index.

Perhaps, the largest source of error in ground-based photometry is from the air
mass extinction, which is described in \citet{Landolt07}. In summary, the air
mass correction is derived from 4--6 stars in standard fields that are observed
every few hours on every night at air-mass values similar to the program stars.
The extinction for any program star is interpolated in time from the set of
standard star determinations. Small air mass differences between standards and
program stars are accounted by differences in the secant of the zenith angles.
Rapid changes in atmospheric transmission will cause errors of order 0.01 mag,
but the expectation is that those errors are random and will be reduced by
repeated observations of the program stars over many nights. Because the
atmospheric extinction decreases with wavelength from U to I \citep{hayes75},
better agreement between space and ground based fluxes might be expected in the
longer wavelength filters, except that there are strong time-variable absorption
lines due to H$_2$O in the I band. Thus, our focus is on V and R for the
Landolt/CALSPEC comparison.

\subsection{Variability of BD+17$^{\circ}$4708}

Figure~\ref{bd17} shows the individual observations of the Sloan standard
BD+17$^{\circ}$4708 \citep{fukugita96}, which are the basis for the magnitude
and color indicies of BD+17$^{\circ}$4708 in table 4 of \citet{Landoltuom07}.
Linear fits to the data points after JD2446500 in 1986 suggest that star is
variable with an increase in flux of $\sim$8 mmag/yr in bands UBVR over the
1986-91 time period. The
statistical significance of this brightening ranges up to
7$\sigma$, while similar analyses for other program stars produce slopes with
less than $\sim$3$\sigma$ significance. The second most likely variable is
GRW+70$^{\circ}$5824 with a decline in brightness in U and B of 5--6 mmag/yr,
but with only 3$\sigma$ significance; in VRI, the decline is $\sim$3 mmag/yr
with only 2$\sigma$ significance.

\section{DISCUSSION} 

\subsection{Equations}

The HST CALSPEC standard star spectrophotometry $F_{\lambda}$ with units erg
cm$^{-2}$ s$^{-1}$ \AA$^{-1}$ is compared with Landolt photometry using the
technique of synthetic photometry. The mean flux in wavelength units over a
photometric bandpass function R is \begin{equation}\langle F\rangle={\int
F_{\lambda}~\lambda~R~d\lambda \over \int
\lambda~R~d\lambda}\label{fav}\end{equation} \citep{Bohlinetal14, bessell12},
where R is the unitless system transmission or quantum efficiency (QE). A
particular magnitude system is defined relative to a reference flux 
$\langle~F_{o}\rangle$ as, \begin{equation}m_{\lambda}=-2.5~log(\langle
F\rangle/\langle F_{o}\rangle)=-2.5~log(\langle
F\rangle)+ZP,\label{mlam}\end{equation} e.g. for Vega magnitudes, which are
defined as zero at all wavelengths; and $\langle F\rangle=\langle F_{o}\rangle$
would be the CALSPEC flux of Vega ($alpha\_lyr\_stis\_008.fits$). However, the
Johnson magnitudes of Vega are non-zero; and the Johnson zero point (ZP) for
each star is  \begin{equation}ZP=2.5~log(\langle
F_{o}\rangle)=m_{\lambda}+2.5~log(\langle F\rangle),\label{zp}\end{equation}
where $m_{\lambda}$ is the Landolt stellar magnitude on the Johnson system and 
$\langle F\rangle$ is the Equation (\ref{fav}) integral of the CALSPEC fluxes.

\subsection{Comparison of Landolt Photometry with CALSPEC SEDs}

For the comparison of the Landolt UBVRI photometry with the CALSPEC SEDs,
bandpass functions from \citet{cohen03}, \citet{maiz06}, and \citet{bessell12}
are investigated. \citet{bessell12} say that smooth functions must be fit to
their coarsely sampled bandpasses; but fitting splines sampled every  Angstrom
makes less than 0.001 mag difference in the computed synthetic photometry. While
\citet{maiz06} and \citet{bessell12} estimate the actual Johnson-Cousins system
throughput for the filters, the Cohen functions are from measured transmission
of the Landolt filters as multiplied by typical atmospheric transmission.
Because the Cohen bandpass functions are based on measurements and are estimates
for the actual Landolt instrumentation, the color transformations from the
instrumental magnitudes to the Johnson-Cousins system must be added to the
synthetic photometry. Also, the Cohen bandpasses differ from  Ma{\'{\i}z
Apell\'aniz and Bessell and from the QE R in Equation (\ref{fav}), i.e. there is
a wavelength factor $\lambda$ included. Thus, the Cohen bandpass functions must
be divided by $\lambda$ for comparison with  Ma{\'{\i}z Apell\'aniz and Bessell.
The wavelength vectors for all three ground-based systems are converted to
vacuum before multiplication by the CALSPEC vacuum SEDs $F_{\lambda}$ in
Equation (\ref{fav}).

For each band in each system, the ZPs from Equation (\ref{zp}) are computed for
each star; and the weighted mean and rms of the zero points are compiled in
Table 3, where the best agreement is for the  Ma{\'{\i}z Apell\'aniz V with only
0.007 mag of scatter among the 11 non-variable stars. For example,
Figure~\ref{alcfbmv} illustrates the differences between Landolt and CALSPEC for
this best case, where BD+17$^{\circ}$4708 is shown but not included in the rms
scatter. However, the two coolest stars, P177D and P330E, are high in comparison
with the hot stars. This systematic trend could be ameliorated if the bandpass
function is slightly in error.

Figure~\ref{bp} illustrates the three V bandpasses in comparison with the steep
SED for a hot star and a flatter SED for a cooler star. A shift of any bandpass
toward shorter wavelengths decreases the relative synthetic flux ratio of the 
cool to hot star. While the Bessell bandpass produces similar results to the
Ma{\'{\i}z Apell\'aniz results in Figure~\ref{bp}, the Cohen results are in the
opposite sense, where a shift of the Cohen bandpass toward longer wavelengths
would be required to improve the cool/hot star agreement. Only Cohen and Bessell
provide bandpass functions for all of the five Landolt bands; and Table 3 shows
that the Cohen rms is generally the worst, despite his valient attempt to derive
the transmissions directly from first principles. Thus, for a uniform result
across all five bands, the Bessell bands are slightly shifted in order to
minimize the rms scatter for the 11 stars. These wavelength shifts for
optimizing the Landolt/CALSPEC comparison are in Table 4 along with the optimum
zero-point reference fluxes $F_{o}$, the zero points in mag, and rms scatter,
while Figure~\ref{alcf-vopt} illustrates the improvement over
Figure~\ref{alcfbmv} for the V band. There are only 12 CALSPEC stars with
Landolt photometry and complete STIS coverage of the V band. However, there are
an additional four stars with complete coverage in R and I.
Figure~\ref{alcf-ropt} also shows sub-percent agreement between the actual and
synthetic photometry in the shifted Bessell R band, where the atmospheric
extinction and time-variable absorption lines are minimal and comparable to the
V band. In V, only AGK+81$^{\circ}$~266 shows as much as a 2$\sigma$ difference
between the actual and synthetic CALSPEC photometry; and similarly in R, there
are no serious discrepancies from perfect agreement between the two independent
measures of stellar flux, (except for the variable star BD+17$^{\circ}$4708,
which differs by 0.014 mag in both V and R).

While the Bessell bandpasses are referenced to photometry with considerable
weight on SAAO data, \citet{menzies91} found small systematic differences
between SAAO and Landolt photometry as a function of RA. Our shifts of the
Bessell bands may represent actual differences between SAAO and Landolt
bandpasses or, perhaps, just reflect subtle differences between the SAAO and
Landolt representations of the UBVRI system.

Using the optimally shifted Bessell bandpasses for the 11 or 15 non-variable
stars, Table 5 contains the magnitudes and uncertainties of Vega on the
Johnson-Cousins system of UBVRI photometry, where $m(Vega)$ is defined by
Equation (\ref{mlam}) using the version $stis\_008$ CALSPEC flux $F(Vega)$. The
uncertainties are just for the conversion to magnitudes and do not include any
uncertainty in the CALSPEC flux itself. For comparison, Table 5 also includes
UBVRI from \citet{bessell12} and the original measures of \citet{johnson66}. 
Our results agree well with \citet{bessell12}, except in the problematic U  band
where variable atmospheric extinction provides the short wavelength cutoff of
the filter transmission function. The \citet{johnson66} photometry also agrees
within 0.01 mag in B and V but should not be expected to agree for the more
commonly used Cousins R and I bandpasses used here.

As a final check for systematic errors, Figure~\ref{alcfam} shows the same
optimized difference in photometry vs. air mass for the V band. The weighted
linear, least square fit suggests that any systematic error in the airmass
correction is less than 5 mmag. The slope of the fitted line differs
from zero by less than 3$\sigma$, so that our data are consistent with
no error in the airmass correction.

\acknowledgements

AUL wishes to thank the then Diretor of Lowell Observatory, Robert L. Millis,
for generous telescope time assignments for his photometric standard star
program. This aspect of AUL's observational program has been supported by NSF
grants AST-0503871 and AST 0803158. Primary support for RCB was provided by NASA
through the Space Telescope Science Institute, which is operated by AURA, Inc.,
under NASA contract NAS5-26555. This research made use of the SIMBAD database,
operated at CDS, Strasbourg, France.

\begin{figure}
\centering 
\includegraphics[height=6in]{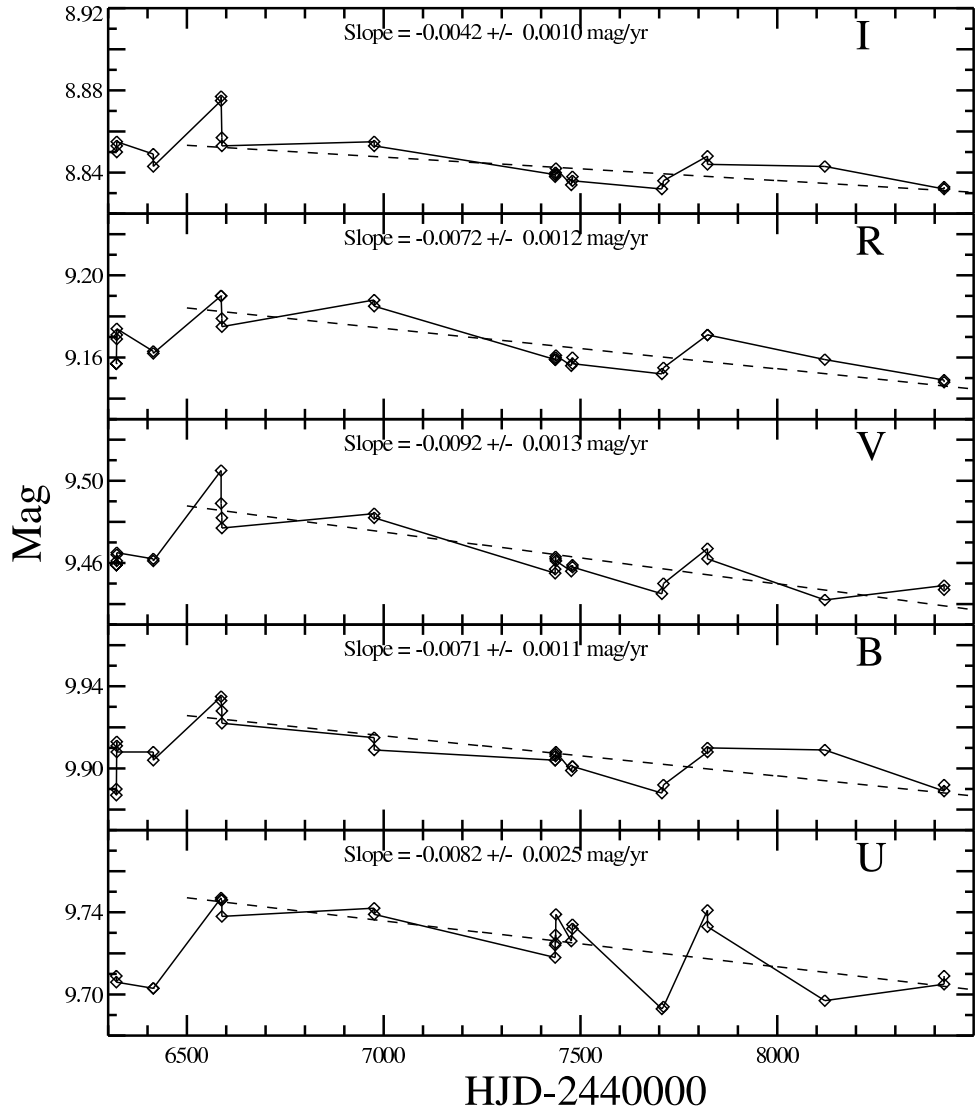}
\caption{\baselineskip=14pt
Variation of the brightness of BD+17$^{\circ}$4708 in various bands.
\label{bd17}}
\end{figure}

\begin{figure}
\centering 
\includegraphics[height=5in]{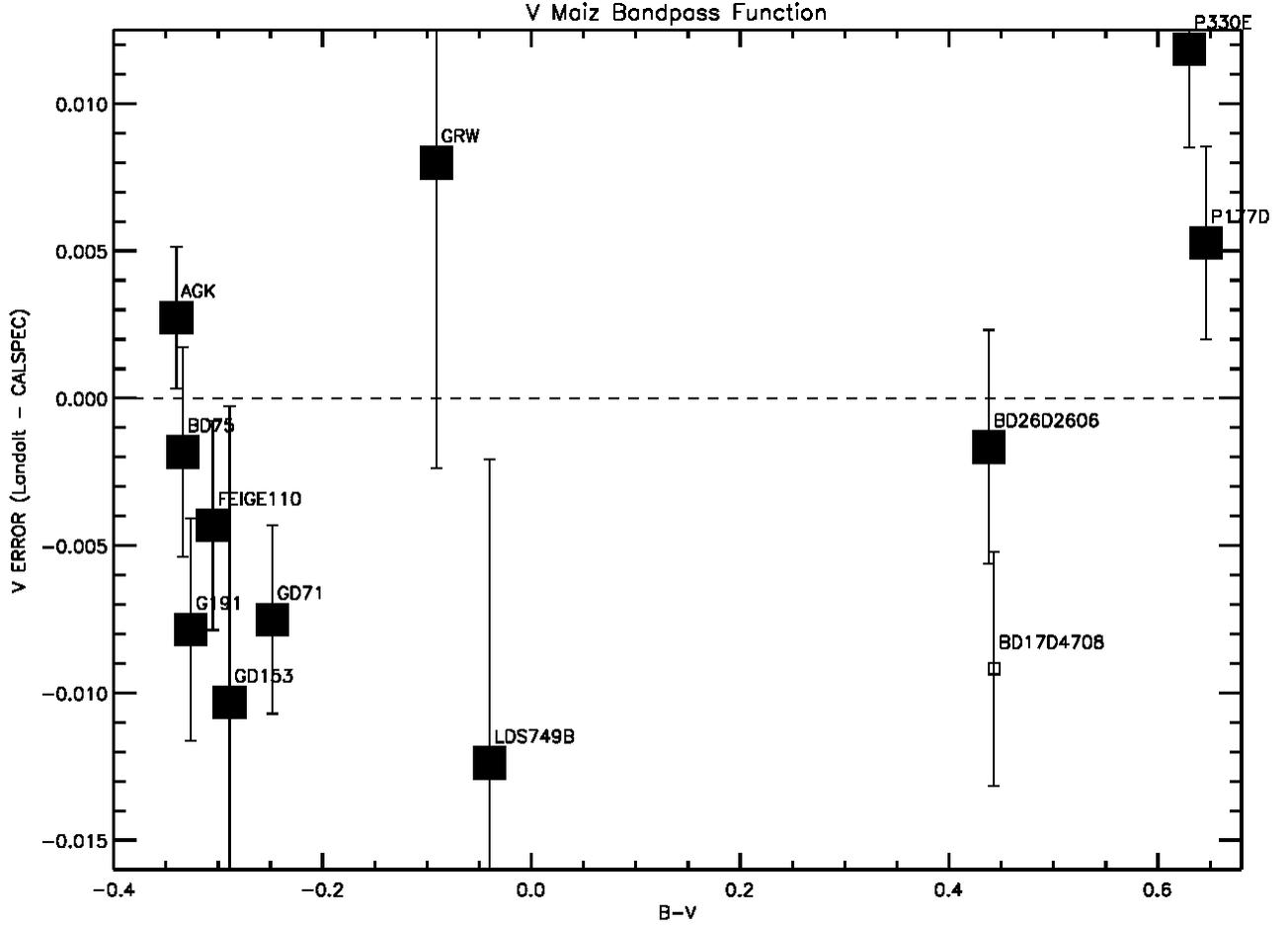}
\caption{\baselineskip=14pt
Comparision of STIS synthetic photometry to Landolt actual photometry in the V
band using the \citet{maiz06} bandpass function for the synthetic photometry.
The poorly observed GRW+70$^{\circ}$5824 and the faint LDS749B have large STIS
uncertainties. GD153 was observed only four times by Landolt.
\label{alcfbmv}} \end{figure}

\begin{figure}
\centering 
\includegraphics[height=5in]{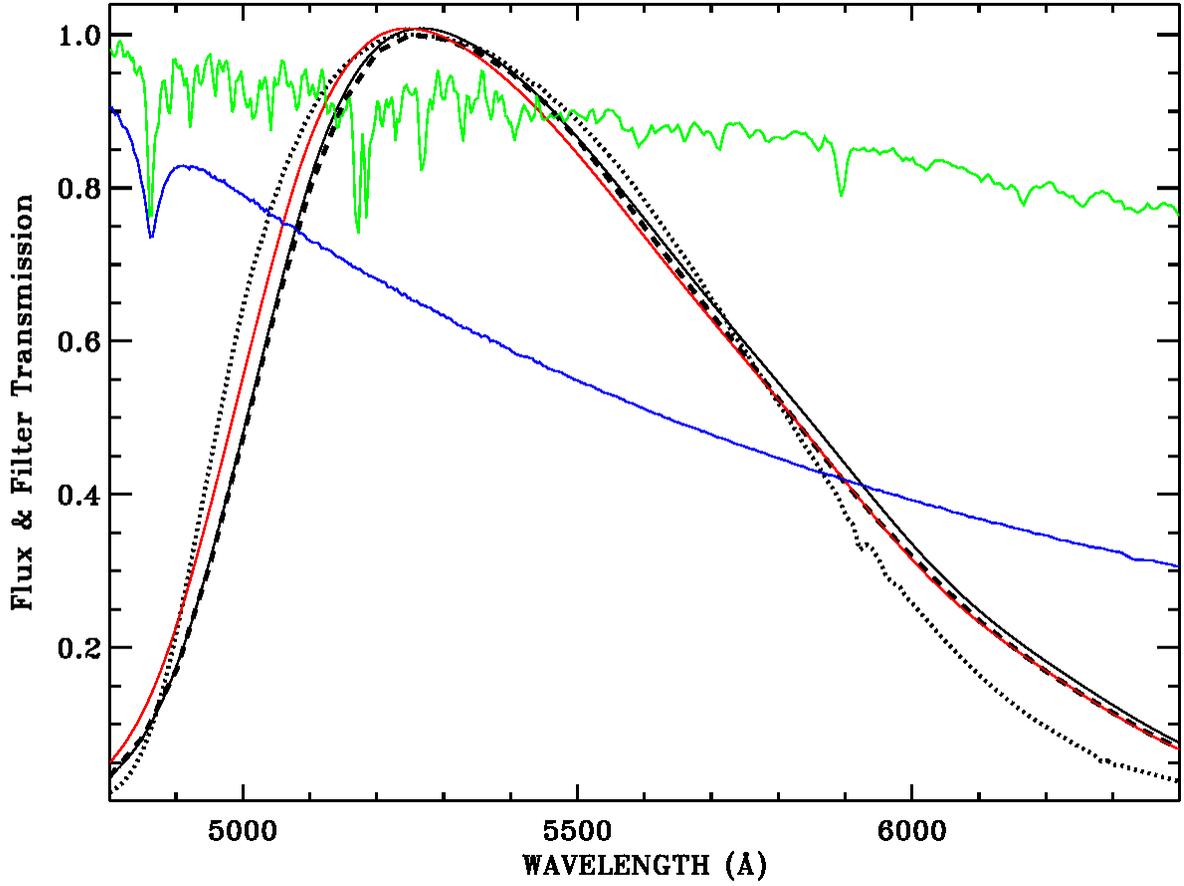}
\caption{\baselineskip=14pt
Normalized V band filter curves from \citet{cohen03}:dots, \citet{maiz06}:dash,
and \citet{bessell12}:solid. The Cohen relative response has been divided by
wavelength. Normalized CALSPEC SEDs are shown for G191B2B:blue and for
P330E:green. The red curve is the Bessell bandpass shifted by -20 \AA~to
minimize tbe Landolt/CALSPEC differences. \label{bp}} \end{figure}

\begin{figure}
\centering 
\includegraphics[height=5in]{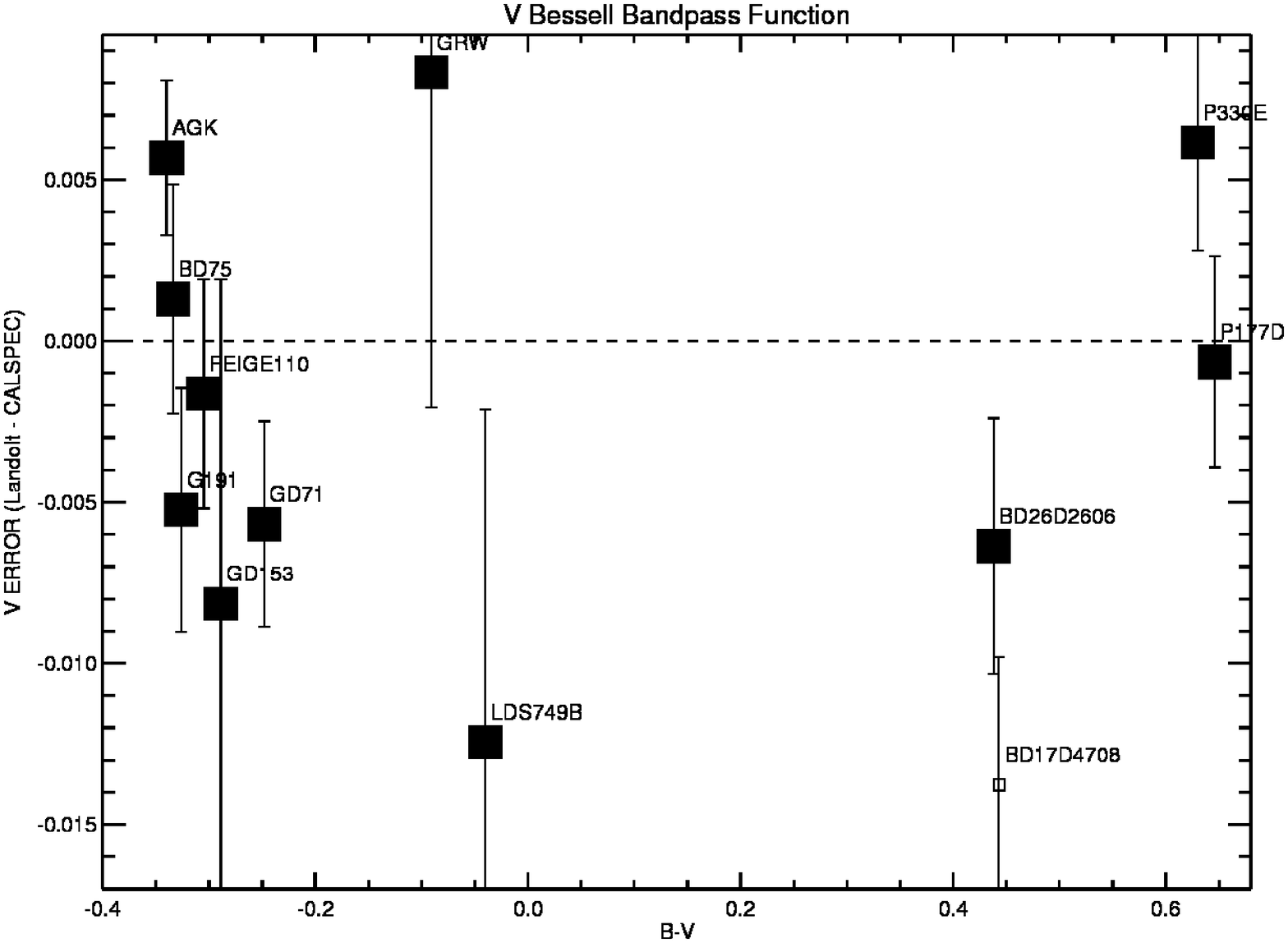}
\caption{\baselineskip=14pt
As in Figure~\ref{alcfbmv} except for the \citet{bessell12} bandpass shifted 
by -20 \AA~to optimize the agreement of the actual and synthetic photometry.
\label{alcf-vopt}} \end{figure}

\begin{figure}
\centering 
\includegraphics[height=5in]{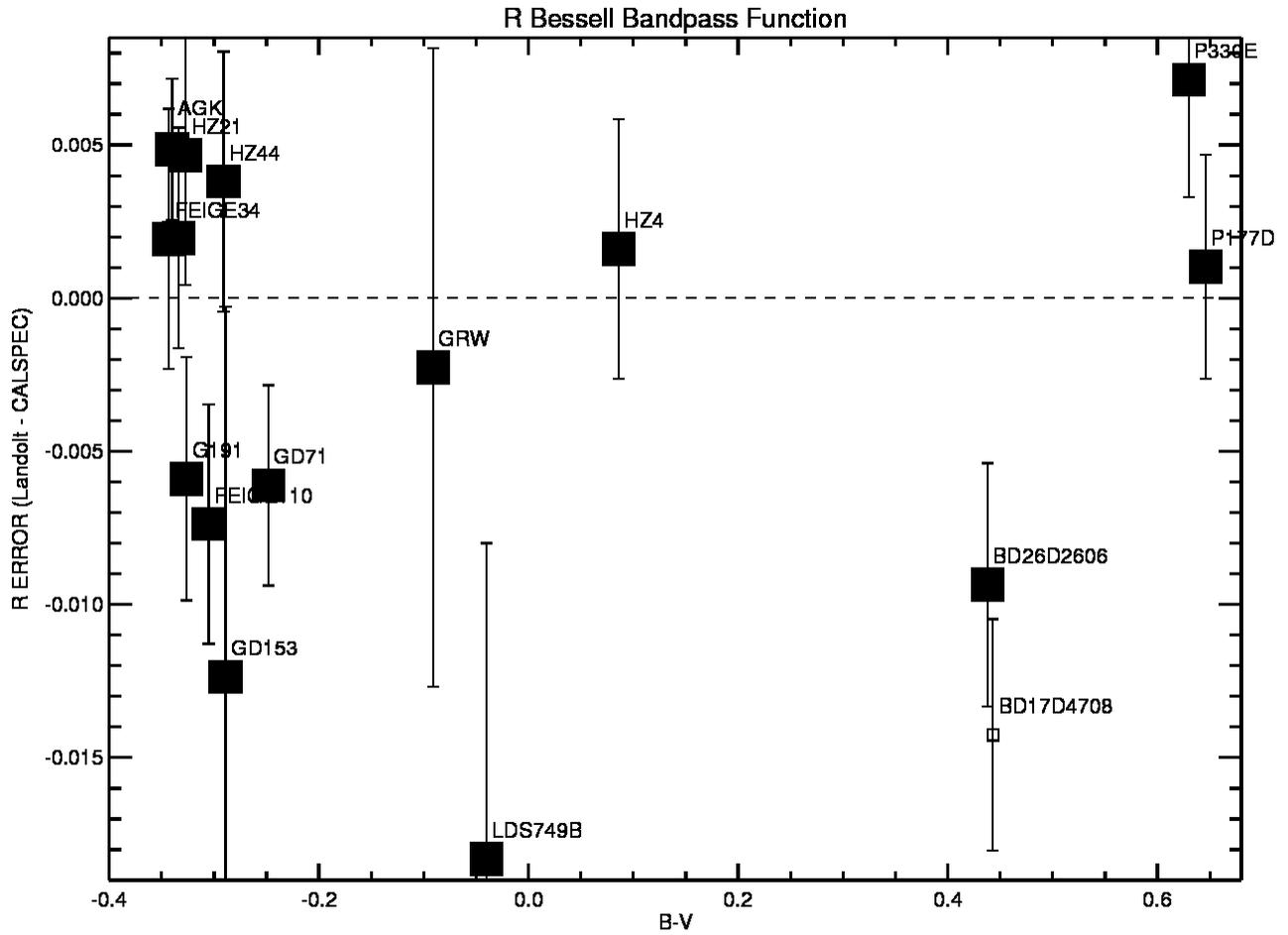}
\caption{\baselineskip=14pt
As in Figure~\ref{alcf-vopt} except for the R band shifted 
by -31 \AA~to optimize the agreement of the actual and synthetic photometry.
\label{alcf-ropt}} \end{figure}

\begin{figure}
\centering 
\includegraphics[height=5in]{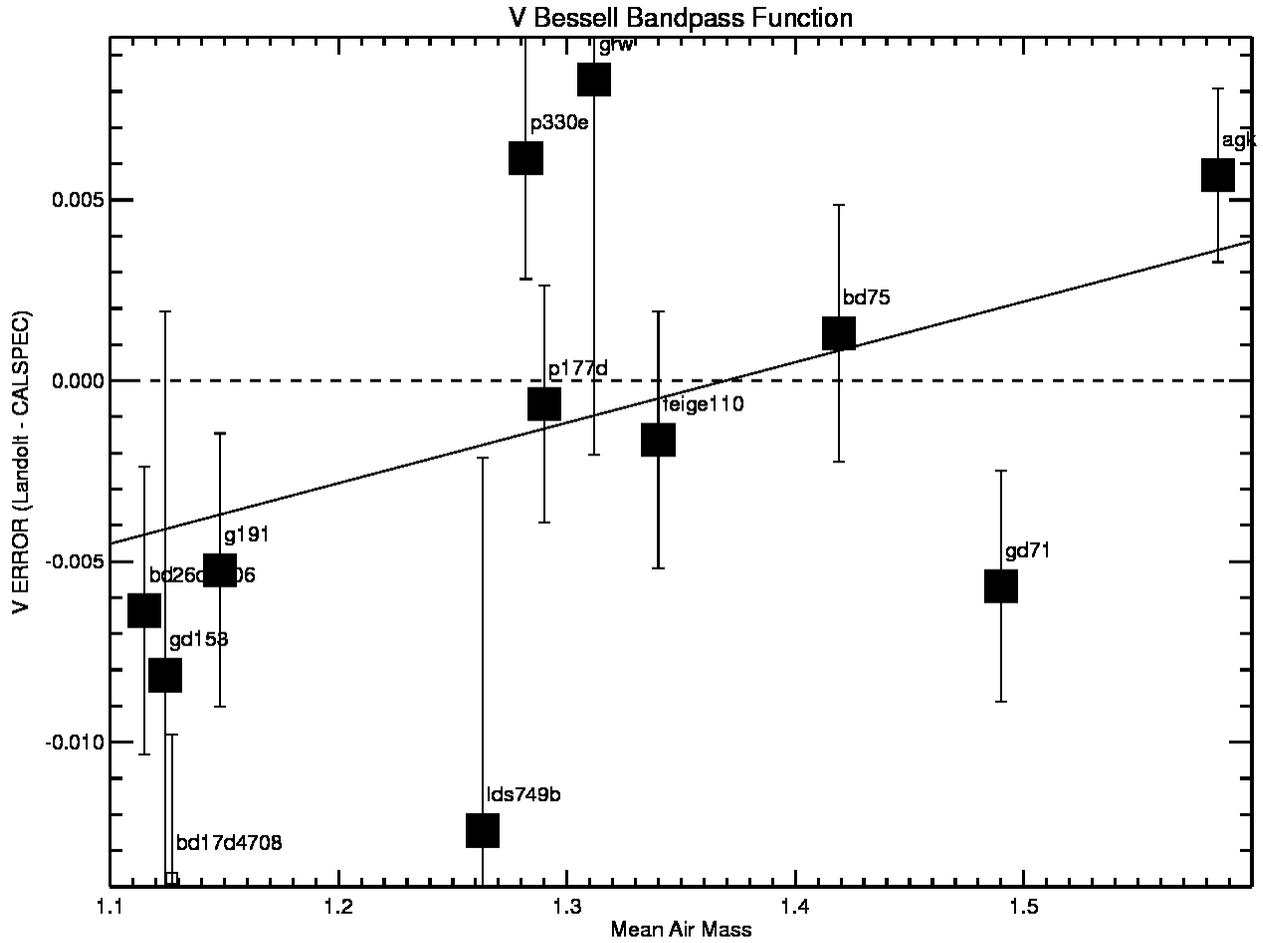}
\caption{\baselineskip=14pt
As in Figure~\ref{alcfbmv}, except plotted vs. air mass and for the
shifted \citet{bessell12} bandpass function. The solid line is the weighted
linear, least-square fit to the 11 filled black squares.
\label{alcfam}} \end{figure}


\begin{deluxetable}{ccccrrrrr}
\tablewidth{0pt}
\tablecolumns{9}
\tablecaption{Average UBVRI Photometry and rms Uncertainty for P177-D and P330-E}
\tablehead{
\colhead{Star} &\colhead{RA (2000)} &\colhead{DEC (2000)}
&\colhead{V} &\colhead{B-V} &\colhead{U-B} &\colhead{V-R} &\colhead{R-I} &\colhead{V-I} }
\startdata
P177-D	&15 59 13.579 &+47 36 41.91 &13.492&+0.646 &+0.156 &+0.364 &+0.371 &+0.737 \\
Unc in Mean &         &             &0.004 &0.009  &0.016  &0.004  &0.010  &0.013  \\
P330-E  &16 31 33.82  &+30 08 46.5  &13.028&+0.630 &+0.070 &+0.362 &+0.362 &+0.726 \\
Unc in Mean &         &             &0.004 &0.006  &0.010  &0.007  &0.007  &0.008  \\
GD153   &12 57 02.337 &+22 01 52.68 &13.349&-0.289 &-1.177 &-0.139 &-0.180 &-0.320 \\
rms     &             &             &0.008 &0.007  &0.010  &0.013  &0.017   &0.005 \\
\enddata
\end{deluxetable}

\begin{deluxetable}{cccrrrrrr}
\tablewidth{0pt}
\tablecolumns{9}
\tablecaption{Individual Observations}
\tablehead{
\colhead{Star} &\colhead{UT Date} &\colhead{HJD}
&\colhead{V} &\colhead{B-V} &\colhead{U-B} &\colhead{V-R} &\colhead{R-I} 
&\colhead{V-I}\\
&\colhead{mmddyy}&}

\startdata
P177-D		&080907	&2454321.70656  &13.489  &+0.638  &+0.155  &+0.363  &+0.373  &+0.738\\
P177-D		&080907	&2454321.71088  &13.492  &+0.645  &+0.162  &+0.367  &+0.371  &+0.741\\
P177-D		&081007	&2454322.71172  &13.497  &+0.667  &+0.124  &+0.362  &+0.363  &+0.728\\
P177-D		&081007	&2454322.71626  &13.500  &+0.634  &+0.169  &+0.358  &+0.361  &+0.722\\
P177-D 		&081107	&2454323.71285  &13.486  &+0.637  &+0.172  &+0.357  &+0.356  &+0.715\\
P177-D		&081107	&2454323.71722  &13.489  &+0.649  &+0.157  &+0.358  &+0.368  &+0.729\\
P177-D		&090308 &2454712.68389  &13.495  &+0.652  &+0.161  &+0.367  &+0.394  &+0.763\\
P177-D		&090308 &2454712.68863  &13.491  &+0.641  &+0.177  &+0.367  &+0.374  &+0.744\\
P177-D		&090408 &2454713.66759  &13.490  &+0.652  &+0.143  &+0.369  &+0.376  &+0.746\\
P177-D		&090408 &2454713.67186  &13.490  &+0.650  &+0.139  &+0.369  &+0.370  &+0.740\\
  Avg		&	&		&13.492  &+0.646  &+0.156  &+0.364  &+0.371  &+0.737\\
  rms		&	&		& 0.004  & 0.009  & 0.016  & 0.004  & 0.010  & 0.013 \\
  Unc in Mean 	&	&n = 10 	& 0.0013 & 0.0028 & 0.0051 & 0.0013 & 0.0032 & 0.0041\\
\\
P330-E		&080907  &2454321.71661  &13.036  &+0.622  &+0.066  &+0.370  &+0.348  &+0.721\\
P330-E		&080907  &2454321.72013  &13.025  &+0.624  &+0.071  &+0.364  &+0.360  &+0.726\\
P330-E		&081007  &2454322.72194  &13.030  &+0.634  &+0.075  &+0.352  &+0.371  &+0.725\\
P330-E		&081007  &2454322.72544  &13.031  &+0.633  &+0.077  &+0.354  &+0.358  &+0.715\\
P330-E		&081107  &2454323.72212  &13.023  &+0.623  &+0.077  &+0.365  &+0.356  &+0.724\\
P330-E		&081107  &2454323.72562  &13.028  &+0.622  &+0.070  &+0.362  &+0.371  &+0.735\\
P330-E		&090308  &2454712.69367  &13.024  &+0.639  &+0.078  &+0.354  &+0.363  &+0.720\\
P330-E		&090308  &2454712.69747  &13.030  &+0.638  &+0.076  &+0.362  &+0.356  &+0.720\\
P330-E		&090408  &2454713.67720  &13.021  &+0.635  &+0.054  &+0.365  &+0.369  &+0.736\\
P330-E		&090408  &2454713.68053  &13.032  &+0.629  &+0.050  &+0.373  &+0.366  &+0.740\\
  Avg	     	&	 &	 	 &13.028  &+0.630  &+0.070  &+0.362  &+0.362  &+0.726\\
  rms	     	&	 &	 	 & 0.004  & 0.006  & 0.010  & 0.007  & 0.007  & 0.008\\
  Unc in Mean	&	 &n = 10	 & 0.0013 & 0.0019 & 0.0032 & 0.0022 & 0.0022 & 0.0025\\
\enddata
\end{deluxetable}

\begin{deluxetable}{ccrrrrr}
\tablewidth{0pt}
\tablecolumns{7}
\tablecaption{Average Zero Points and rms Scatter for Three Bandpass Functions}
\tablehead{
\colhead{Ref.} &\colhead{ZP or rms} 
&\colhead{U} &\colhead{B} &\colhead{V} &\colhead{R} &\colhead{I} }
\startdata
\citet{cohen03}   &ZP\tablenotemark{a} &4.483 &6.831 &3.772 &2.263 &1.121\\
		    &ZP &-20.871 &-20.414 &-21.059 &-21.614 &-22.376\\
                   &rms &0.065 &0.040 &0.011 &0.008 &0.018 \\
\citet{maiz06}    &ZP\tablenotemark{a} &4.238 &6.333 &3.674 & & \\
                   &ZP &-20.932 &-20.496 &-21.087 & & \\
                   & rms &0.015 &0.011 &0.007 & & \\
\citet{bessell12} &ZP\tablenotemark{a}  &4.212 &6.314 &3.659 
&2.198 &1.169 \\
                    &ZP  &-20.939 &-20.499 &-21.092 &-21.645 &-22.330 \\
                   & rms &0.017 &0.013 &0.008 &0.008 &0.010 \\
\enddata
\tablenotetext{a}{$\langle F_{o}\rangle~(10^{-9}$ erg cm$^{-2}$ s$^{-1}$
			 \AA$^{-1})$. 
Other ZP and rms values are mag units.}
\end{deluxetable}

\begin{deluxetable}{ccccc}
\tablewidth{0pt}
\tablecolumns{5}
\tablecaption{Shifts, ZPs, and rms Scatter for the Optimized Bessell Bandpass Functions}
\tablehead{
\colhead{Band} &\colhead{Shift} &\colhead{ZP=$\langle F_{o}\rangle$} 
&\colhead{ZP} &\colhead{rms} \\
&\colhead{(\AA)} &\colhead{$(10^{-9}$ erg cm$^{-2}$ s$^{-1}$\AA$^{-1})$} 
&\colhead{(mag)} &\colhead{(mag)}}
\startdata
U   &-8   &4.232  &-20.934  &.0160   \\
B   &-20  &6.396  &-20.485  &.0073  \\
V   &-20  &3.700  &-21.079  &.0054  \\
R   &-31  &2.236  &-21.626  &.0059  \\
I   &-27  &1.184  &-22.317  &.0096  \\
\enddata
\end{deluxetable}

\begin{deluxetable}{ccccrrrr}
\tablewidth{0pt}
\tablecolumns{8} 
\tablecaption{Magnitudes and Uncertainty for Vega using the Shifted 
Bessell Bandpass Functions}
\tablehead{
\colhead{Star} &\colhead{RA (2000)} &\colhead{DEC (2000)}
&\colhead{U} &\colhead{B} &\colhead{V} &\colhead{R} &\colhead{I} }
\startdata
Vega    &18 36 56.336 &+38 47 01.28 &0.064 &0.020 &0.028 &0.033 &0.029 \\
rms     &             &             &0.016 &0.007 &0.005 &0.006 &0.010 \\
Unc in Mean 	&	&           &0.007 &0.004 &0.003 &0.003 &0.004 \\
\\
Bessell    &            &           &0.041 &0.023 &0.027 &0.027 &0.028 \\
Johnson    &            &           &0.03  &0.03  &0.03  &0.07\tablenotemark{a}
			&0.10\tablenotemark{a} \\
\enddata
\tablenotetext{a}{These Johnson R and I values are not expected to agree 
with the shorter wavelength Cousins R and I values in the rest
of the Table.}
\end{deluxetable}

\clearpage

\bibliographystyle{apj}
\bibliography{../../../pub/paper-bibliog}

\end{document}